\documentclass[11pt]{article}
\usepackage{amssymb}
\usepackage{epsfig}

\newcommand{\BE}{\begin{equation}}
\newcommand{\EE}{\end{equation}}
\newcommand{\BA}{\begin{eqnarray}}
\newcommand{\EA}{\end{eqnarray}}

\begin{document}

\title{Evidence for an anisotropy of the speed of light } 
\author{
C. M. L. de Arag\~ao$^1$\thanks{Cristiane.Aragao@ct.infn.it}~, 
M. Consoli$^1$\thanks{Maurizio.Consoli@ct.infn.it}~
and A. Grillo$^2$\thanks{agrillo@dmfci.unict.it}~ \\
$^1$  {\it \small Istituto Nazionale di Fisica Nucleare, 
Sezione di Catania} \\
{\it \small  Via Santa Sofia 64, 95123 Catania, Italy } \\
\\ $^2$ {\it \small DMFCI, Facolt\`a di Ingegneria, Universit\`a di Catania} \\
{\it \small Viale Andrea Doria 6, 95126 Catania, Italy}} 
\maketitle
\thispagestyle{empty}
\date{}

\begin{abstract}
By comparing with the most recent experimental results, we 
point out the model dependence of the present bounds on the 
anisotropy of the speed of light. In fact, by replacing the CMB with
a class of preferred frames that can better account for the experimental 
data, one obtains values of the RMS anisotropy parameter 
$(1/2 -\beta + \delta)$ 
that are one order of magnitude larger than the presently quoted ones. 
The resulting non-zero anisotropy can be understood starting from the
observation that the speed of light in the Earth's gravitational 
field is {\it not} the basic parameter $c=1$ entering Lorentz 
transformations. In this sense, light can propagate isotropically 
only in one `preferred' frame.
\end{abstract}


\pagebreak

\addtocounter{page}{-1}

\section{Introduction}

The idea of a preferred reference frame dates back to the origin of the
Theory of Relativity, i.e. to the basic differences between Einstein's 
Special Relativity and the Lorentzian point of view. Today the former 
approach is generally accepted and with it the interpretation of 
the relativistic effects in terms of the relative motion between any 
pair S' and S'' of observers. However, in spite of the deep conceptual
differences, as emphasized by Bell \cite{bell,brown}, it is not so 
simple to distinguish experimentally between the two alternatives. 
In fact, relativistic effects might be interpreted, equally well, 
as arising from the {\it individual} motions of each observer 
with respect to a preferred frame $\Sigma$. In this case, the basic 
Lorentz transformations would be associated with the velocity 
parameters $\beta'=v'/c$ and $\beta''=v''/c$, $v'$ and $v''$ 
being the velocities of S' and S'' with respect to $\Sigma$
(we restrict for simplicity to the one-dimensional case). 
The equivalence of the two formulations is then a simple consequence
of the basic group property of Lorentz transformations where the 
relation between S' and S'' is also a Lorentz transformation 
with relative velocity parameter 
\BE
\label{zero}
       \beta_{\rm rel}= {{ \beta' - \beta''}\over{1- \beta'\beta''}}.
\EE
It would be possible to distinguish the two formulations if the individual
parameters $\beta'$ and $\beta''$ could be experimentally determined through
e\-ther-drift experiments. At the same time, it is clear that this possibility 
crucially depends on having a speed of light in the vacuum, say $c_\gamma$, 
that is {\it not} the same parameter $c\equiv 1$ entering Lorentz 
transformations. In this case, light would be seen to propagate isotropically 
with a speed $c_\gamma\neq c\equiv 1$ in $\Sigma$. However, in any other 
moving frame S' the speed of light would exhibit a non-trivial angular 
dependence $c_\gamma(\theta)$ induced
by the Lorentz transformation connecting S' to $\Sigma$.

Notice that, once the space-time transformations are taken to be
Lorentz transformations, the basic isotropy and homogeneity of space and time 
(assumed to be valid in $\Sigma$) hold true in any other moving frame S' as
well. Thus for S' the length of a rod, at rest in S', does not depend 
on its orientation. This means that for S' the length $L$ of a 
resonating cavity (at rest in S') cannot depend on the angle 
$\theta$ between the cavity axis and the velocity of S' with 
respect to $\Sigma$. Therefore, in the general relation between 
the cavity frequency $\nu=\nu(\theta)$, the cavity length 
$L=L(\theta)$ and the two-way speed of light $\bar{c}_\gamma(\theta)$ 
\BE
       \nu(\theta)\sim {{ \bar{c}_\gamma(\theta) }\over{ L(\theta) }}
\EE
one can take $L(\theta)=L=$ constant if Lorentz transformations are valid.
In this way, the relative frequency shift of two orthogonal optical resonators
\begin{eqnarray}
\label{basic0}
       {{ \delta \nu (\theta) }\over{\nu_0}} \equiv
       {{ \nu(\pi/2 +\theta)- \nu (\theta)} 
\over {\nu_0 }}
       = {{\bar{c}_\gamma(\pi/2 +\theta)- \bar{c}_\gamma
       (\theta)} \over {\langle \bar{c}_\gamma \rangle }} 
\end{eqnarray}
provides a direct measurement of the anisotropy of the speed of light in the
rest frame S' of the apparatus.

As a possible scenario for a value $c_\gamma \neq c\equiv 1$ one can 
consider, for instance, models with extra space-time dimensions 
\cite{extra}. These represent an interesting approach toward a 
consistent quantum theory of gravity and predict tipically 
a speed of gravity $c_g \neq c$. This leads, in the 4D effective 
theory, to a version of relativity where there is a preferred 
frame $\Sigma$, the one associated with the isotropic value of $c_g$. 
At the same time, through the coupling to gravitons, the induced 
Lorentz-violations \cite{lorentz} will extend to the other sectors 
of the theory. Namely, through the effect of graviton loops
the photon energy spectrum becomes
\BE
\label{one}
       E_\gamma(|{\bf{p}}|)=c_\gamma|{\bf{p}}| ,
\EE
where $c_\gamma$ differs from the basic parameter $c\equiv 1$ 
entering Lorentz transformations.

Therefore, assuming the correct space-time transformations
to be Lorentz transformations, the i\-so\-tro\-pic relation Eq.~(\ref{one}) 
cannot hold true in more than one frame. In this sense, the vacuum 
can be viewed as a physical medium with a non-trivial refractive index 
${\cal N}_{\rm vacuum}\equiv {{1}\over{c_\gamma}} \neq 1$. Thus, 
if light is seen to propagate isotropically by the observer in 
$\Sigma$, on the Earth there would be a small anisotropy of the two-way 
speed of light
\BE
\label{anisotropy}
       {{\delta \bar{c}_\gamma}\over{ \bar{c}_\gamma }}\sim
       ({\cal N}_{\rm vacuum}- 1)
       {{v^2_{\rm earth}}\over{c^2}} ,
\EE
$v_{\rm earth}$ being the Earth's velocity with respect to $\Sigma$.

We emphasize that the idea of extra space-time dimensions is just an 
example of theoretical framework that can produce a vacuum value 
$c_\gamma \neq c\equiv 1$. In fact, the same conclusion 
applies equally well to the more conventional case 
of a background gravitational field
\cite{pagano}. To better appreciate this remark, 
let us observe that the
universality of free fall, at the base of the Equivalence Principle and of
local Lorentz invariance, guarantees the identity of the local speed of
light with the basic parameter $c\equiv 1$ entering 
Lorentz transformations.

However, for an observer sitting at rest on the Earth's surface, 
this identification does not account for the Earth's
gravitational field which
is equivalent to an effective refractive index. In fact, 
from the Earth's gravitational potential 
\BE
\label{two}
       \varphi =- {{G_N M_{\rm earth}}\over{c^2 R_{\rm earth} }}
       \sim -7\cdot 10^{-10}
\EE
and the weak-field isotropic form of the metric \cite{weinberg}
\BE
\label{three}
       ds^2= (1+ 2\varphi) dt^2 - (1-2\varphi)(dx^2 +dy^2 +dz^2)
\EE
one obtains an energy as in Eq.~(\ref{one}) with a value
\BE
\label{four}
       c_\gamma= {{1}\over{{\cal N}_{\rm vacuum} }}\sim 1+ 2\varphi .
\EE
Again, since $c_\gamma\neq c\equiv 1$, Eq.~(\ref{one}) cannot be valid 
in more than one frame.

The aim of this paper is to explore the observable consequences 
of such a scenario where $c_\gamma \neq c\equiv 1$ by comparing 
with the ether-drift experiments and, in particular, with the 
new generation where vacuum cryogenic optical resonators 
are maintained under active rotation.
The existence of a preferred frame $\Sigma$ should produce
periodic modulations of the signal as those associated with 
the typical angular frequency defined by the Earth's rotation.

Comparing with the results of two recent experiments 
\cite{schiller,peters}, we shall show that the present interpretation 
of the data is not unambiguous but strongly depends on restricting 
the hypothetical preferred frame to coincide with the CMB. 
For this reason, we shall report model-independent relations that 
can be used to restrict from experiment the class of possible 
Earth's cosmic motions. At the same time, such a model-independent analysis 
of the present 
data provides substantially different indications on the anisotropy of the
speed of light.

\section{General formalism}

Within the framework outlined in the Introduction, we shall follow 
the authors of Ref.~\cite{burgess} by introducing  a set
of effective Minkowski tensors $\hat{\eta}(i)_{\mu\nu}$
\BE
\label{scheme}
       \hat{\eta}(i)_{\mu\nu}= \eta_{\mu\nu} - \kappa_i v_\mu v_\nu .
\EE
Here $\eta_{\mu\nu}=$diag(-1,1,1,1), $v_\mu$ is the 
4-velocity of S' with respect to the hypothetical preferred 
frame $\Sigma$ while $\kappa_i$ represent generalized
Fresnel's drag coefficients for particles of type $i$.
In the context of the models with extra space-time dimensions 
considered in Ref.~\cite{burgess} they originate from the 
interactions of the various particles with the gravitons. 
In general, Eqs.~(\ref{scheme}) represent a convenient
framework to parameterize the dependence of the results on the 
motion of the observer with respect to a preferred frame.

In this way, the energy-momentum relation in a given frame S' 
can be expressed as
\BE
       p^\mu p^\nu \hat{\eta}(i)_{\mu\nu} + m^2(i)= 0 .
\EE
For photons this becomes
\BE
\label{masshell}
       p^\mu p^\nu \hat{\eta}(\gamma)_{\mu\nu}= 0 ,
\EE
with $\hat{\eta}(\gamma)_{\mu\nu}= \eta_{\mu\nu} - \kappa_\gamma 
v_\mu v_\nu$ and with a photon energy that, in the S' frame, 
depends on the direction between the photon momentum 
and the S' velocity  ${\bf{v}}$ with respect to $\Sigma$.

To obtain the photon energy spectrum we shall follow Jauch 
and Watson \cite{jauch} who also derived Eq.~(\ref{masshell})
working out the quantization of the electromagnetic field in 
a moving medium. They noticed that the procedure introduces 
unavoidably a preferred frame, the one where the photon energy 
does not depend on the direction of propagation, and which is 
``usually taken as the system for which the medium is at rest". 
However, such an identification reflects the point of view of Special 
Relativity with no preferred frame. Therefore, we shall adapt their 
results to our case where the photon energy does not depend 
on the angle in some frame $\Sigma$. In this way, in a moving
frame S', we get the radiation field Hamiltonian 
\BE
       H_0=\sum_{r=1,2}\int d^3{\bf{p}}\left[\hat{n}_r({\bf{p}}) +
       {{1}\over{2}}\right] E(|{\bf{p}}|, \theta) ,
\EE
where $\hat{n}_r({\bf{p}})$ is the photon number operator and
\BE
\label{watson}
       E(| {\bf{p}}| , \theta)= {{ \kappa_\gamma v_0 \zeta
       + \sqrt{ |{\bf{p}}|^2(1+\kappa_\gamma v^2_0) -
       \kappa_\gamma \zeta^2 }}\over{ 1 + \kappa_\gamma v^2_0}}
\EE
with
\BE
       \zeta={\bf{p}}\cdot{\bf{v}}= |{\bf{p}}||{\bf{v}}|\cos\theta ,
\EE
$\theta\equiv\theta_{\rm lab}$ being the angle defined, in the S' frame, 
between the photon momentum and the S' velocity ${\bf{v}}$ with 
respect to $\Sigma$. Notice that only one of the two roots 
of Eq.~(\ref{masshell}) appears and the energy is not positive definite in 
connection with the critical velocity $1/\sqrt{1+\kappa_\gamma}$ 
defined by the occurrence of the Cherenkov radiation.

Using the above relation, the one-way speed of light in the S' frame 
depends on $\theta$ (we replace $v=|{\bf{v}}|$ and $v^2_0=1 +v^2$) 
\begin{eqnarray}
      {{E(| {\bf{p}}| , \theta)}\over{|{\bf{p}}|}}=
       c_\gamma(\theta)= 
       {{ \kappa_\gamma v \sqrt{1+v^2} \cos\theta +
       \sqrt{ 1+ \kappa_\gamma+ \kappa_\gamma v^2 \sin^2\theta} }
       \over{1+ \kappa_\gamma(1+v^2)}} .
\end{eqnarray}
This is different from the $v=0$ result, in the $\Sigma$ frame, where the 
energy does not depend on the angle
\BE
\label{esigma}
       {{E^{(\Sigma)}(| {\bf{p}}|)}\over{|{\bf{p}}|}}=
       c_\gamma= {{1}\over{ {\cal N}_{\rm vacuum} }} 
\EE
and, as in Eq.~(\ref{one}), the speed of light is simply 
rescaled by the inverse of the vacuum refractive index 
\BE
\label{root}
       {\cal N}_{\rm vacuum}=\sqrt{1+ \kappa_\gamma} .
\EE
Working to ${\cal O}(\kappa_\gamma)$ and ${\cal O}(v^2)$, 
one finds in the S' frame
\BE
\label{oneway}
       c_\gamma(\theta)= {{1+ \kappa_\gamma v \cos\theta -
       {{\kappa_\gamma}\over{2}} v^2(1+\cos^2\theta) }\over
       {\sqrt{ 1+\kappa_\gamma} }} .
\EE
This expression differs from Eq. (6) of Ref.~\cite{pla}, for the 
replacement $\cos\theta \to -\cos\theta$ and for the relativistic 
aberration of the angles. In Ref.~\cite{pla}, in fact, the 
one-way speed of light in the S' frame was parameterized in terms 
of the angle $\theta\equiv\theta_\Sigma$, between the velocity of S'
and the direction of propagation of light, as defined in the $\Sigma$ frame. 
In this way, starting from Eq.~(\ref{oneway}), replacing $\cos\theta \to -
\cos\theta$ and using the aberration relation
\BE
       \cos (\theta_{\rm lab})={{-v + \cos\theta_\Sigma}\over
       {1-v\cos\theta_\Sigma}}
\EE
one re-obtains Eq. (6) of Ref.~\cite{pla} in terms of $\theta=\theta_\Sigma$.

Further, using Eq.~(\ref{oneway}), the two-way speed of light (in terms of
$\theta=\theta_{\rm lab}$) is
\begin{eqnarray}
\label{twoway}
       \bar{c}_\gamma(\theta) =
       {{ 2  c_\gamma(\theta) c_\gamma(\pi + \theta) }\over{
       c_\gamma(\theta) + c_\gamma(\pi + \theta) }} 
       \sim 1-\left[\kappa_\gamma -
       {{\kappa_\gamma}\over{2}} \sin^2\theta \right]v^2 .
\end{eqnarray}

Now, re-introducing, for sake of clarity, the speed of light 
entering Lorentz transformations, $c=2.997..\cdot10^{10}$ cm/s, 
one can define the RMS \cite{robertson,mansouri} parameter 
$(1/2 -\beta +\delta)$. This is used to parameterize the 
anisotropy of the speed of light in the vacuum, through the relation 
\BE
\label{rms}
       {{\bar{c}_\gamma(\pi/2 +\theta)- \bar{c}_\gamma (\theta)} \over
       {\langle \bar{c}_\gamma \rangle }} \sim
       (1/2-\beta +\delta)
       {{v^2 }\over{c^2}} \cos(2\theta)
\EE
so that one can relate $\kappa_\gamma$ to $(1/2 -\beta +\delta)$ 
through
\BE
\label{rmsgamma}
       (1/2 -\beta +\delta)= {{\kappa_\gamma}\over{2}} .
\EE
These results can be easily applied to the propagation of photons in a 
background gravitational field, such as on the Earth's surface. In this 
case, if there were a preferred frame, the S' energy would {\it not} 
be given by Eq.~(\ref{one}) but would rather be given by Eq.~(\ref{watson}) 
with a value of $\kappa_\gamma$ obtained from Eqs. ~(\ref{four}) 
and (\ref{root})
\BE
\label{kappagamma}
       \kappa_\gamma={\cal N}^2_{\rm vacuum} - 1 \sim 28\cdot 10^{-10} .
\EE
This corresponds to a RMS parameter
\BE
\label{kapparms}
       (1/2-\beta +\delta)\sim
       {\cal N}_{\rm vacuum} - 1 \sim 14\cdot 10^{-10}
\EE
and should produce an anisotropy of the two-way speed of light 
in modern ether-drift experiments.

\section{Cosmic motions and ether-drift experiments}

In modern ether-drift experiments, one measures the relative 
frequency shift $\delta \nu$ of two vacuum cryogenic 
optical resonators under the Earth's rotation \cite{muller} 
or upon active rotations of the apparatus \cite{schiller,peters}. 
If there is a preferred frame $\Sigma$, using Eqs.~(\ref{twoway}) and
(\ref{rms}), the frequency shift of two orthogonal optical resonators
to ${\cal O}({{v^2}\over{c^2}})$ can be expressed as
\BE
\label{basic}
       {{\delta \nu (\theta) }\over{\nu_0}}=
       {{\bar{c}_\gamma(\pi/2 +\theta)- \bar{c}_\gamma
       (\theta)} \over {\langle \bar{c}_\gamma \rangle }} =
       A \cos(2\theta) ,
\EE
where $\theta=0$ indicates the direction of the ether-drift and 
the amplitude of the signal is given by
\BE
\label{amplitude}
       A= (1/2 -\beta +\delta) {{v^2 }\over{c^2}} ,
\EE
$v$ denoting the projection of the Earth's velocity with respect 
to $\Sigma$ in the plane of the interferometer.

To address the problem in a model-in\-de\-pen\-dent way, let us 
introduce the time-dependent amplitude of the ether-drift effect
\BE
\label{amplitude1}
       A(t)= v^2(t) X 
\EE
in terms of the time-dependent Earth's velocity and of the 
normalization of the experiment $X$. The main point is that 
the relative variations of the signal depend only on the kinematic
details of the given cosmic motion and, as such, can be predicted 
independently of the knowledge of $X$. To describe the variations 
of $v(t)$, we shall use the expressions given by Nassau and Morse 
\cite{nassau}. These have the advantage of being fully 
model-independent and extremely easy to handle. 
Their simplicity depends on the introduction of a cosmic Earth's velocity 
\BE
\label{vincl}
       {\bf{V}}={\bf{V}}_{\rm sun} +
       {\bf{v}}_{\rm orb}
\EE
that, in addition to the genuine cosmic motion of the solar system 
defined by ${\bf{V}}_{\rm sun}$, includes the effect of the Earth's 
orbital motion around the Sun described by ${\bf{v}}_{\rm orb}$. 
To a very good approximation, ${\bf{V}}$ can be taken to be
constant within short observation periods of 2-3 days. Therefore, 
by introducing the latitude of the laboratory $\phi$, the right
ascension $\tilde{\Phi}$ and the declination $\tilde{\Theta}$ 
associated with the vector ${\bf{V}}$, the magnitude of the
Earth's velocity in the plane of the interferometer is defined by 
the two equations \cite{nassau}
\BE
       \cos z(t)= \sin\tilde{\Theta}\sin \phi + \cos\tilde{\Theta}
       \cos\phi \cos(\lambda)
\EE
and 
\BE
\label{vearth}
       v(t)=V \sin z(t) ,
\EE
$z=z(t)$ being the zenithal distance of ${\bf{V}}$. Here, we have 
introduced the time $\lambda\equiv \tau -\tau_o-\tilde{\Phi}$, where
$\tau=\omega_{\rm sid}t$ is the sidereal time of the observation
in degrees and $\omega_{\rm sid}\sim {{2\pi}\over{23^{h}56'}}$. 
Also, $\tau_o$ is an offset that, in general, has to be introduced 
to compare with the definition of sidereal time adopted in 
Refs.~\cite{schiller,peters}.

Now, operation of the interferometer provides the minimum and maximum 
daily values of the amplitude and, as such, the values $v_{\rm min}$ 
and $v_{\rm max}$ corresponding to $|\cos(\lambda)|=1$. In this way, 
using the above relations one can determine the pair of values
$(\tilde{\Phi}_i,\tilde{\Theta}_i)$, $i=1,2,..n$, for each of the $n$ 
short periods of observations taken during the year, and thus plot 
the direction of the vectors ${\bf{V}}_i$ on the celestial sphere.

Actually, since the ether-drift is a second-har\-mon\-ic effect in 
the rotation angle of the interferometer, a single observation is 
unable to distinguish the pair $(\tilde{\Phi}_i,\tilde{\Theta}_i)$ 
from the pair $(\tilde{\Phi}_i +180^o,$ $-\tilde{\Theta}_i)$. 
Only repeating the observations in different ep\-ochs of the year 
one can resolve the ambiguity. Any meaningful ether-drift, in fact, 
has to correspond to pairs $(\tilde{\Phi}_i,\tilde{\Theta}_i)$ 
lying on an `aberration circle', defined by the Earth's orbital 
motion, whose center $(\Phi,\Theta)$ defines the right ascension 
and the declination of the genuine cosmic motion of the solar 
system associated with ${\bf{V}}_{\rm sun}$. If such a 
consistency is found, using the triangle law, one can finally 
determine the magnitude $|{\bf{V}}_{\rm sun}|$ starting 
from the known values of $(\tilde{\Phi}_i,\tilde{\Theta}_i)$, 
$(\Phi,\Theta)$ and the value $|{\bf{v}}_{\rm orb}|\sim $ 30 km/s. 
In simple terms, 
for $|{\bf{v}}_{\rm orb}|\ll |{\bf{V}}_{\rm sun}|$, the opening angle
$\Delta \varphi$ 
defined by the aberration circle can be approximated as
$\Delta\varphi \sim
  {{ |{\bf{v}}_{\rm orb}| }\over{ |{\bf{V}}_{\rm sun}| }}$.

We emphasize that the kinematical solution of the Earth's cosmic motion, 
as obtained from the basic pairs of values 
$(\tilde{\Phi}_i,\tilde{\Theta}_i)$,
only depends on the {\it relative} magnitude of the ether drift 
effect, namely on the ratio ${{ v_{\rm min} }\over{v_{\rm max} }}$, 
in the various periods. As such, it is insensitive to any possible 
theoretical and/or experimental uncertainty that can affect 
multiplicatively the {\it absolute} normalization of the signal.

For instance, suppose one measures a relative frequency 
shift $\delta \nu/\nu\sim 10^{-15}$. Assuming a value 
$(1/2-\beta+\delta)\sim 10\cdot  10^{-10}$ in Eq.~(\ref{amplitude}), 
this would be interpreted in terms of a velocity $v \sim 300$ km/s. 
Within Galileian relativity, where one predicts the same 
expressions by simply replacing $(1/2-\beta+\delta) \to 1/2$, 
the same frequency shift would be interpreted in terms of
a velocity $v \sim 14$ m/s. Nevertheless, from the {\it relative} 
variations of the ether-drift effect one would deduce the same pairs 
$(\tilde{\Phi}_i,\tilde{\Theta}_i)$ and, as such, exactly the same 
type of cosmic motion. Just for this reason, Miller's determinations 
with this method, namely \cite{miller} ${\bf{V}}_{\rm sun}$ 
$\sim 210$ km/s, $\Phi \sim 74^o$ and $\Theta \sim -70^o$, should be 
taken seriously.

We are aware that Miller's observations have been considered spurious 
by the authors of Ref. ~\cite{shankland} as partly due to statistical 
fluctuations and/or thermal fluctuations. However, to a closer look 
(see the discussion given in Ref.~\cite{cimento}) the arguments
of Ref.~\cite{shankland} are not so solid as they appear by reading 
the abstract of that paper. Moreover, Miller's solution is {\it doubly}
internally consistent since the aberration circle due to the Earth's 
orbital motion was obtained in two different and independent ways
(see Fig. 23 of Ref.~\cite{miller}). In fact, one can determine the 
basic pairs $(\tilde{\Phi}_i,\tilde{\Theta}_i)$ either using the 
daily variations of the magnitude of the ether-drift effect or using
the daily variations of its apparent direction $\theta_0(t)$ 
(the `azimuth') in the plane of the interferometer. Since
the two methods were found to give consistent results, in addition 
to the standard choice of preferred frame represented by the CMB, 
we shall also compare with the cosmic motion deduced by Miller.

Replacing Eq.~(\ref{vearth}) into Eq.~(\ref{amplitude}) and
adopting a notation of the type introduced in Ref.~\cite{mewes}, 
we can express the theoretical amplitude of the signal as 
\begin{eqnarray}
\label{amorse}
       A(t) = A_0 +
       A_1\sin\tau +A_2 \cos\tau 
        +  A_3\sin(2\tau) +A_4 \cos(2\tau) ,
\end{eqnarray}
where ($\chi=90^o-\phi$)
\begin{eqnarray}
\label{a0}
       A_0 = (1/2-\beta+\delta) {{ V^2 }\over{c^2}}
       \left(1- \sin^2\tilde{\Theta}\cos^2\chi 
       - {{1}\over{2}} \cos^2\tilde{\Theta}\sin^2\chi \right) ,
\end{eqnarray}
\BE
\label{a1}
       A_1=-{{1}\over{2}}(1/2-\beta+\delta)
       {{ V^2 }\over{c^2}} \sin 2\tilde{\Theta}
       \sin(\tilde{\Phi} +\tau_o) \sin 2\chi ,
\EE
\BE
\label{a2}
       A_2=-{{1}\over{2}}(1/2-\beta+\delta)
       {{ V^2 }\over{c^2}} \sin 2\tilde{\Theta}
       \cos(\tilde{\Phi} +\tau_o) \sin 2\chi ,
\EE
\BE
\label{a3}
       A_3=-{{1}\over{2}}(1/2-\beta+\delta)
       {{ V^2 }\over{c^2}} \cos^2 \tilde{\Theta}
       \sin[2(\tilde{\Phi} +\tau_o)] \sin^2 \chi  , 
\EE
%
%
\BE
\label{a4}
       A_4=-{{1}\over{2}}(1/2-\beta+\delta)
       {{ V^2 }\over{c^2}} \cos^2 \tilde{\Theta}
       \cos[2(\tilde{\Phi} +\tau_o)] \sin^2 \chi  .
\EE
Recall that $V$, $\tilde{\Theta}$ and $\tilde{\Phi}$ indicate 
respectively the magnitude, the declination and the right 
ascension of the velocity defined in Eq.~(\ref{vincl}). As such, 
they change during the year.

To compare with the experiments of Refs.~\cite{schiller,peters}, however, 
it will be more convenient to re-write Eq.~(\ref{basic}) in the form
of Ref.~\cite{schiller} where the frequency shift at a given time 
$t$ is expressed as
\BE
\label{basic2}
       {{\delta \nu [\theta(t)]}\over{\nu_0}} =
       \hat{B}(t)\sin 2\theta(t) +
       \hat{C}(t)\cos 2\theta(t)
\EE
$\theta(t)$ being the angle of rotation of the apparatus,
$\hat{B}(t)\equiv 2B(t)$ and $\hat{C}(t)\equiv 2C(t)$ so that
one finds an experimental amplitude
\BE
\label{expampli}
       A(t)= \sqrt { \hat{B}^2(t) + \hat{C}^2(t) } .
\EE
In this case, using now Eqs. (21-22) of Ref.~\cite{nassau}, one finds
\begin{eqnarray}
\label{amorse1}
       \hat{C}(t) = \hat{C}_0 +
       \hat{C}_1\sin\tau +\hat{C}_2 \cos\tau
        + \hat{C}_3\sin(2\tau) +\hat{C}_4 \cos(2\tau) ,
\end{eqnarray}
where
\begin{eqnarray}
\label{C0}
       \hat{C}_0 = (1/2-\beta+\delta) {{ V^2 }\over{c^2}}
       {{\sin^2\chi}\over{2}} (3\cos^2\tilde{\Theta} -2) ,
\end{eqnarray}
\BE
\label{C1}
       \hat{C}_1= {{1}\over{2}}(1/2-\beta+\delta) {{ V^2 }\over{c^2}}
       \sin 2\tilde{\Theta} \sin(\tilde{\Phi} +\tau_o) \sin 2\chi ,
\EE
\BE
\label{C2}
       \hat{C}_2={{1}\over{2}}(1/2-\beta+\delta)
       {{ V^2 }\over{c^2}} \sin 2\tilde{\Theta}
       \cos(\tilde{\Phi} +\tau_o) \sin 2\chi ,
\EE
\begin{eqnarray}
\label{C3}
       \hat{C}_3 = -{{1}\over{2}}(1/2-\beta+\delta)
       {{ V^2 }\over{c^2}} \cos^2 \tilde{\Theta}
       \sin[2(\tilde{\Phi} +\tau_o)]
        (1+ \cos^2\chi) ,
\end{eqnarray}
\begin{eqnarray}
\label{C4}
       \hat{C}_4 = -{{1}\over{2}}(1/2-\beta+\delta)
       {{ V^2 }\over{c^2}} \cos^2 \tilde{\Theta}
       \cos[2(\tilde{\Phi} +\tau_o)]
       (1+ \cos^2\chi) .
\end{eqnarray}
Analogously, we find
\begin{eqnarray}
\label{amorse2}
       \hat{B}(t) = \hat{B}_1\sin\tau
       +\hat{B}_2 \cos\tau 
        + \hat{B}_3\sin(2\tau) +\hat{B}_4 \cos(2\tau) ,
\end{eqnarray}
with $\hat{B}_1=-\hat{C}_2/\cos\chi$, $\hat{B}_2=\hat{C}_1/\cos\chi$, 
$\hat{B}_3= -{{2\cos\chi}\over{1+\cos^2\chi}}\hat{C}_4$ and
$\hat{B}_4= {{2\cos\chi}\over{1+\cos^2\chi}}\hat{C}_3$. These 
expressions are in full agreement with the results reported in 
Table I of Ref.~\cite{peters} (for $\tau_o=180^o$).

\section{Comparison with the experimental results}

Let us now compare the above theoretical predictions
with the experimental results of Ref. ~\cite{schiller} (referring
to the short period February 6th- February 8th 2005) 
and with those of Ref.~\cite{peters} (summarizing the observations 
from December 2004 to April 2005). To this end, 
we shall concentrate on 
the observed time modulation of the signal, i.e. on the
quantities $\hat{C}_1,\hat{C}_2,\hat{C}_3,\hat{C}_4$ (and on 
their $\hat{B}$-counterparts). In fact, the average values 
$\langle {\hat C} \rangle $ and $\langle {\hat B} \rangle$ 
are most likely affected by spurious systematic effects as thermal 
drift (see in particular the discussion in Ref.~\cite{schiller} 
and the corresponding one in Ref.~\cite{peters} about the
non-zero value of $B_0$, there called $S_0$).

We shall report in Table 1 the experimental values for the combinations
\BE
\label{sid}
       \hat{C} (\omega_{\rm sid})\equiv \sqrt{\hat{C}^2_1
       + \hat{C}^2_2}
\EE
and
\BE
\label{2sid}
       \hat{C} (2\omega_{\rm sid})\equiv \sqrt{\hat{C}^2_3
       + \hat{C}^2_4}
\EE
and for their $\hat{B}$-counterparts. This is useful to 
reduce the model dependence in the data analysis. In this way, in fact, 
the right ascension $\tilde{\Phi}$ and the offset $\tau_o$ 
drop out in the theoretical predictions that will
only depend on $\tilde{\Theta}$, $V$ and $(1/2-\beta+\delta)$.

At the same time, since Ref.~\cite{peters} provides data that 
have been averaged over various periods of the year, we shall 
parameterize the predictions in terms of the average declination 
$\langle\tilde{\Theta}\rangle = \Theta$ and of the average 
velocity $\langle V \rangle = V_{\rm sun}$ obtaining the 
relations Ref.~\cite{mewes}
\BE
\label{sid1}
       \hat{C} (\omega_{\rm sid})\sim {{1}\over{2}}
       (1/2 -\beta +\delta) {{ V^2_{\rm sun} }\over{c^2}}
       |\sin 2\Theta| \sin 2\chi
\EE
and
\BE
\label{sid2}
       \hat{C} (2\omega_{\rm sid})\sim {{1}\over{2}}
       (1/2 -\beta +\delta){{V^2_{\rm sun}}\over{c^2}}\cos^2\Theta
       (1+ \cos^2\chi) .
\EE
The corresponding $\hat{B}$-quantities are also given by
$\hat{B} (\omega_{\rm sid})= \hat{C} (\omega_{\rm sid})/\cos\chi$ and 
$\hat{B} (2\omega_{\rm sid})= {{2\cos\chi}\over{1+\cos^2\chi}}$ 
$\hat{C} (2\omega_{\rm sid})$. Notice that $V_{\rm sun}$ and 
$(1/2-\beta+\delta)$ are completely correlated in each single
measurement. There\-fore, an unambiguous extraction of the RMS
parameter in ether-drift experiments cannot be done without 
a preliminary determination of the cosmic velocity. In turn, 
as explained in Sect. 3, this depends on the possibility to
use the triangle laws after observation of the Earth's aberration circle. 
Since the 
present data just cover a small portion of the Earth's orbit, we 
shall first compare with the experimental data of Refs.
\cite{schiller,peters} assuming in the above relations the fixed 
values $V_{\rm sun}\sim 370$ km/s and $\Theta\sim -6^o$ 
that correspond to the Earth's motion relatively to the CMB. 
In this way, we obtain the values of the RMS parameter 
reported in the third column of Table 1 (the colatitude 
of the laboratory has been taken $\chi \sim 39^o$ for
Ref.~\cite{schiller} and $\chi \sim 38^o$ for Ref.~\cite{peters}).

\begin{table*}
\caption{The experimental data 
and the values of the RMS parameter obtained by 
constraining the hypothetical preferred frame to coincide with the CMB. }
\begin{center}
\begin{tabular}{cll}
\hline\hline
Experiment&Observable &$(1/2-\beta+\delta)$\\
\hline
Ref.\cite{schiller}& $\hat{C} (\omega_{\rm sid})=(11 \pm 2)\cdot 10^{-16}$ &
$(71\pm 13)\cdot 10^{-10}$ \\
Ref.\cite{peters}& $\hat{C} (\omega_{\rm sid})=(3.0 \pm 2.4)\cdot 10^{-16}$ &
$(20\pm 16 )\cdot 10^{-10}$ \\
Ref.\cite{peters}& $\hat{B} (\omega_{\rm sid})=(8.4 \pm 4.4)\cdot 10^{-16}$ &
$(43\pm 23 )\cdot 10^{-10}$ \\
Ref.\cite{schiller}& $\hat{C} (2\omega_{\rm sid})=(1 \pm 2)\cdot 10^{-16}$ &
$(0.8\pm 1.6)\cdot 10^{-10}$ \\
Ref.\cite{peters}& $\hat{C} (2\omega_{\rm sid})=(2.3 \pm 3.4)\cdot 10^{-16}$ &
$(1.9\pm 2.8 )\cdot 10^{-10}$ \\
Ref.\cite{peters}& $\hat{B} (2\omega_{\rm sid})=(4.8 \pm 2.6)\cdot 10^{-16}$ &
$(4.0\pm 2.2 )\cdot 10^{-10}$ \\
\hline\hline
\end{tabular}
\end{center}
\end{table*}

As one can see, the experimental data at the frequency 
$\omega=\omega_{\rm sid}$ produce sistematically higher 
estimates of the RMS parameter. In fact, averaging only the 
determinations from observables at $\omega=\omega_{\rm sid}$ gives
$(1/2-\beta+\delta)\sim (45\pm 10) \cdot 10^{-10}$. This should 
be compared with the value $(1/2-\beta+\delta)\sim (2\pm 2) 
\cdot 10^{-10}$ obtained from the data at $\omega=2\omega_{\rm sid}$. 
On the other hand, averaging all determinations gives again 
the smaller value $(1/2-\beta+\delta)\sim (2\pm 2) \cdot 10^{-10}$ 
but the chi-square of the mean is unacceptably large.

It is not difficult to understand the reason for such a discrepancy. 
It originates from the inadequacy of the CMB type of motion to describe 
some basic features of the data. In fact, using the theoretical prediction 
\BE
\label{ratio}
       R\equiv  {{ \hat{C} (2\omega_{\rm sid}) }\over{
       \hat{C}(\omega_{\rm sid}) }} \sim
       {{0.82}\over{|\tan \Theta|}} ,
\EE
for the latitude of the two experiments, one would expect a value
\BE
\label{ratio1}
       R_{\rm CMB} \sim  7.8 .
\EE
This is very far from the precise experimental value of Ref.~\cite{schiller}
\BE
\label{expdecl}
       R_{\rm EXP}(\mbox{Ref.\cite{schiller}}) \sim 0.09^{+0.18}_{-0.09}
\EE
that would rather require $|\Theta|\sim {83^o}^{+7^o}_{-12^o}$.

Analogously, introducing the other ratio
\BE
\label{ratioprime}
       R'\equiv  {{ \hat{B} (2\omega_{\rm sid}) }\over{
       \hat{B}(\omega_{\rm sid}) }} \sim
       {{0.62}\over{|\tan \Theta|}} , 
\EE
the theoretical prediction for $|\Theta|\sim 6^o$ 
\BE
\label{ratio1prime}
       R'_{\rm CMB} \sim  5.9
\EE
does not agree with the experimental result from Ref.~\cite{peters}
\BE
\label{expdeclprime}
       R'_{\rm EXP}({\rm Ref}.\cite{peters}) \sim 0.57^{+0.63}_{-0.31}
\EE
that would rather require $|\Theta|\sim (47^o\pm 21^o)$ \cite{note}.

Starting from these observations, one can check the RMS parameters that 
would rather be obtained from the various observables replacing the CMB 
with a class of preferred frames in better consistency with
Eq.~(\ref{expdecl}). For instance, replacing $|\Theta|\sim 6^o$ with the 
value $|\Theta|\sim 70^o$ (as in Miller's solution) obtained
by averaging the $\Theta$-values from
Eqs.~(\ref{expdecl}) and (\ref{expdeclprime}), and 
leaving out $V_{\rm sun}$ as a free parameter, the results change 
substantially. In fact, the values of $(1/2-\beta+\delta)$ from
the observables at $\omega=\omega_{\rm sid}$ are now decreased by a factor
${{\sin (12^o)}\over{\sin (140^o)}}\sim 0.32$ while the values of 
$(1/2-\beta+\delta)$ from the observables at $\omega=2\omega_{\rm sid}$ 
are now increased by a factor ${{ \cos^2(6^o) }\over { \cos^2(70^o)}} 
\sim 8.4$. These new results are shown in Table 2 where we have 
also introduced the additional rescaling $k$, defined through the 
relation $\sqrt{k}={{370~ {\rm km/s} }\over{V_{\rm sun} }}$, to treat
$V_{\rm sun}$ as a free parameter.

\begin{table*}
\caption{The experimental data and the values of the RMS parameter 
obtained by replacing the CMB with a class of preferred frames with 
declination $|\Theta|=70^o$. $k$ accounts for the possible values of 
the Sun velocity as discussed in the text.}
\begin{center}
\begin{tabular}{cll}
\hline\hline
Experiment&Observable &$(1/2-\beta+\delta)$\\
\hline
Ref.\cite{schiller}& $\hat{C} (\omega_{\rm sid})=(11 \pm 2)\cdot 10^{-16}$ &
$k(23\pm 4)\cdot 10^{-10}$ \\
Ref.\cite{peters}& $\hat{C} (\omega_{\rm sid})=(3.0 \pm 2.4)\cdot 10^{-16}$ &
$k(6\pm 5 )\cdot 10^{-10}$ \\
Ref.\cite{peters}& $\hat{B} (\omega_{\rm sid})=(8.4 \pm 4.4)\cdot 10^{-16}$ &
$k(14\pm 7 )\cdot 10^{-10}$ \\
Ref.\cite{schiller}& $\hat{C} (2\omega_{\rm sid})=(1 \pm 2)\cdot 10^{-16}$ &
$k(7\pm 14)\cdot 10^{-10}$ \\
Ref.\cite{peters}& $\hat{C} (2\omega_{\rm sid})=(2.3 \pm 3.4)\cdot 10^{-16}$ &
$k(16\pm 24 )\cdot 10^{-10}$ \\
Ref.\cite{peters}& $\hat{B} (2\omega_{\rm sid})=(4.8 \pm 2.6)\cdot 10^{-16}$ &
$k(34\pm 19 )\cdot 10^{-10}$ \\
\hline\hline
\end{tabular}
\end{center}
\end{table*}

As one can see, independently of $k$, the values of the RMS parameter 
obtained from all observables are now in good consistency with 
each other. In fact, the average value is $(1/2-\beta+\delta)
\sim k(17\pm 3) \cdot 10^{-10}$ with a good chi-square per degree of 
freedom. Notice also that, for $k\sim 1$ there would also be 
a good consistency with the prediction $(1/2-\beta+\delta)\sim 
14\cdot 10^{-10}$ expected on the base of Eq.~(\ref{kapparms}).

\section{Summary and outlook}

In this paper we have explored some phenomenological consequences of assuming
the existence of a preferred frame. This scenario, that on the one hand
leads us back to the old Lorentzian version of relativity, is also
favoured by present models with extra space-time dimensions where
the interactions with the gravitons change the vacuum into a physical 
medium with a non-trivial refractive index where the speed of light 
\BE
       c_\gamma={{1}\over{ {\cal N}_{\rm vacuum} }}
\EE
differs from the basic parameter $c\equiv 1$ entering Lorentz 
transformations. In this case, where light can propagate 
isotropically in just one (preferred) frame $\Sigma$, in any 
other frame there would be an anisotropy that could be detected 
through ether-drift experiments.

Our main point is that there is another simpler mechanism accounting 
for $c_\gamma \neq c\equiv 1$ : the Earth's background gravitational 
field. This introduces an effective vacuum refractive index
\BE
       {\cal N}_{\rm vacuum} \sim 1 - 2\varphi ,
\EE
with 
\BE
       \varphi =- {{G_N M_{\rm earth}}\over{c^2 R_{\rm earth} }} \sim
       -7\cdot 10^{-10}
\EE
and corresponds to a RMS parameter
\BE
\label{earth}
       (1/2-\beta +\delta)\sim
       {\cal N}_{\rm vacuum} - 1 \sim 14\cdot 10^{-10} .
\EE
Thus, if there were a preferred frame $\Sigma$ where light 
is seen isotropic, one should be able to detect some 
effect with the new generation of precise ether-drift 
experiments using rotating cryogenic optical resonators. 
In particular, one should look for periodic modulations 
of the signal that might be associated with the Earth's 
rotation and its orbital motion around the Sun.

When comparing with the experimental results of Ref.~\cite{schiller,peters} 
we can draw the following conclusions. Assuming the preferred frame to
coincide with the CMB ($V_{\rm sun}\sim 370$ km/s and an average 
declination $\Theta \sim -6^o$) the observables at the Earth's 
rotation frequency $\omega=\omega_{\rm sid}$ are consistent 
with rather a large value of the RMS parameter 
$(1/2-\beta+\delta) \sim (45 \pm 10)\cdot 10^{-10}$ (see Table 1). 
At the same time, the signal at $\omega=2\omega_{\rm sid}$ 
is comparably weaker yielding the much smaller value
$(1/2-\beta+\delta) \sim (2 \pm 2)\cdot 10^{-10}$.

In our opinion, both estimates are likely affected by a systematic
uncertainty of theoretical nature. In fact, as explained in Sect. 4, 
the average declination favoured by the experimental data is 
far from the value $|\Theta|\sim 6^o$ defined by the Earth's 
motion relatively to the CMB. For this reason, and since the physical 
nature of the preferred frame is unknown, one should check the stability 
of $(1/2-\beta+\delta)$ against some other type of cosmic motion. 
In this case, adopting an average declination in better consistency 
with the data, say $|\Theta|\sim 70^o$, and an arbitrary value of the
Sun velocity, as embodied in the rescaling
$\sqrt{k}={{370~{\rm km/s}}\over{V_{\rm sun} }}$, 
one obtains good consistency among all observables (see Table 2) with
an average 
RMS parameter $(1/2-\beta+\delta) \sim k(17 \pm 3)\cdot 10^{-10}$.
For $k$ of order unity, this is consistent with the prediction 
in Eq.~(\ref{earth}).

On the other hand, eliminating 
such $k-$de\-pen\-dence of the RMS parameter requires a preliminary 
determination of $V_{\rm sun}$. As explained in Sect. 3, this can only be
done after observing the effect of
the Earth's orbital motion through the relation 
$\Delta\varphi \sim
  {{ 30~{\rm km/s} }\over{ |{\bf{V}}_{\rm sun}| }}$, 
$\Delta \varphi$ being the opening angle of the
`aberration circle' induced by the Earth's revolution around the Sun. 
In turn, this requires more data, covering larger parts of the Earth's 
orbit, that exhibit seasonal 
modulations of the signal. For instance, Miller's observations were
indicating a variation of the
declination from $|\tilde{\Theta}|\sim 77^o$ in 
February-April to $|\tilde{\Theta}|\sim 62^o$ in 
August-Sep\-tem\-ber. This should correspond to a $\sim +70\%$ increase of 
the daily variations \cite{consoli} in the two periods. 
Such kind of confirmations would represent clean 
experimental evidence for the existence of a preferred frame, 
a result with far-reaching implications for both particle 
physics and cosmology.

\vskip 20 pt

\centerline{\bf{Acknowledgements}}
M. C. thanks N. Cabibbo and P. M. Stevenson for 
useful discussions. C.M.L. de A. thanks the Italian Ministero degli
Affari Esteri (MAE) for financial support.

\vskip 30 pt
\end{document}